     
\font\twelverm=cmr10 scaled 1200    \font\twelvei=cmmi10 scaled 1200
\font\twelvesy=cmsy10 scaled 1200   \font\twelveex=cmex10 scaled 1200
\font\twelvebf=cmbx10 scaled 1200   \font\twelvesl=cmsl10 scaled 1200
\font\twelvett=cmtt10 scaled 1200   \font\twelveit=cmti10 scaled 1200
\font\twelvesc=cmcsc10 scaled 1200  
\skewchar\twelvei='177   \skewchar\twelvesy='60
     
     
\def\twelvepoint{\normalbaselineskip=12.4pt plus 0.1pt minus 0.1pt
  \abovedisplayskip 12.4pt plus 3pt minus 9pt
  \belowdisplayskip 12.4pt plus 3pt minus 9pt
  \abovedisplayshortskip 0pt plus 3pt
  \belowdisplayshortskip 7.2pt plus 3pt minus 4pt
  \smallskipamount=3.6pt plus1.2pt minus1.2pt
  \medskipamount=7.2pt plus2.4pt minus2.4pt
  \bigskipamount=14.4pt plus4.8pt minus4.8pt
  \def\rm{\fam0\twelverm}          \def\it{\fam\itfam\twelveit}%
  \def\sl{\fam\slfam\twelvesl}     \def\bf{\fam\bffam\twelvebf}%
  \def\mit{\fam 1}                 \def\cal{\fam 2}%
  \def\sc{\twelvesc}               \def\tt{\twelvett}
  \def\sf{\twelvesf}
  \textfont0=\twelverm   \scriptfont0=\tenrm   \scriptscriptfont0=\sevenrm
  \textfont1=\twelvei    \scriptfont1=\teni    \scriptscriptfont1=\seveni
  \textfont2=\twelvesy   \scriptfont2=\tensy   \scriptscriptfont2=\sevensy
  \textfont3=\twelveex   \scriptfont3=\twelveex  \scriptscriptfont3=\twelveex
  \textfont\itfam=\twelveit
  \textfont\slfam=\twelvesl
  \textfont\bffam=\twelvebf \scriptfont\bffam=\tenbf
  \scriptscriptfont\bffam=\sevenbf
  \normalbaselines\rm}
     

     
\def\beginlinemode{\endmode
  \begingroup\parskip=0pt \obeylines\def\\{\par}\def\endmode{\par\endgroup}}
\def\beginparmode{\endmode
  \begingroup \def\endmode{\par\endgroup}}
\let\endmode=\par
{\obeylines\gdef\
{}}
\def\singlespace{\baselineskip=\normalbaselineskip}

\def\oneandahalfspace{\baselineskip=\normalbaselineskip
  \multiply\baselineskip by 3 \divide\baselineskip by 2}
\def\doublespace{\baselineskip=\normalbaselineskip \multiply\baselineskip by 2}

\newcount\firstpageno
\firstpageno=2
\footline={\ifnum\pageno<\firstpageno{\hfil}\else{\hfil\twelverm\folio\hfil}\fi}
\def\toppageno{\global\footline={\hfil}\global\headline
  ={\ifnum\pageno<\firstpageno{\hfil}\else{\hfil\twelverm\folio\hfil}\fi}}
\let\rawfootnote=\footnote              
\def\footnote#1#2{{\rm\singlespace\parindent=0pt\parskip=0pt
  \rawfootnote{#1}{#2\hfill\vrule height 0pt depth 6pt width 0pt}}}
\def\raggedcenter{\leftskip=4em plus 12em \rightskip=\leftskip
  \parindent=0pt \parfillskip=0pt \spaceskip=.3333em \xspaceskip=.5em
  \pretolerance=9999 \tolerance=9999
  \hyphenpenalty=9999 \exhyphenpenalty=9999 }
\def\dateline{\rightline{\ifcase\month\or
  January\or February\or March\or April\or May\or June\or
  July\or August\or September\or October\or November\or December\fi
  \space\number\year}}
\def\received{\vskip 3pt plus 0.2fill
 \centerline{\sl (Received\space\ifcase\month\or
  January\or February\or March\or April\or May\or June\or
  July\or August\or September\or October\or November\or December\fi
  \qquad, \number\year)}}
     
     
\hsize=6.5truein
\vsize=8.5truein
\parskip=\medskipamount
\def\\{\cr}
\twelvepoint            
\doublespace            
\overfullrule=0pt       

\def\title                      
  {\null\vskip 3pt plus 0.2fill
   \beginlinemode \doublespace \raggedcenter \bf}
     
\def\author                     
  {\vskip 3pt plus 0.2fill \beginlinemode
   \singlespace \raggedcenter\sc}
     
\def\affil                      
  {\vskip 3pt plus 0.1fill \beginlinemode
   \oneandahalfspace \raggedcenter \sl}
     
\def\abstract                   
  {\vskip 3pt plus 0.3fill \beginparmode
   \singlespace ABSTRACT: }
     
\def\endtopmatter               
  {\endpage                     
   \body}
     
\def\body                       
  {\beginparmode}               
     
\def\head#1{                    
  \goodbreak\vskip 0.5truein    
  {\immediate\write16{#1}
   \raggedcenter \uppercase{#1}\par}
   \nobreak\vskip 0.25truein\nobreak}
     
\def\subhead#1{                 
  \vskip 0.25truein             
  {\raggedcenter {#1} \par}
   \nobreak\vskip 0.25truein\nobreak}
     
\def\beginitems{
\par\medskip\bgroup\def\i##1 {\item{##1}}\def\ii##1 {\itemitem{##1}}
\leftskip=36pt\parskip=0pt}
\def\enditems{\par\egroup}
     
\def\beneathrel#1\under#2{\mathrel{\mathop{#2}\limits_{#1}}}
     
     
\def\references                 
  {\head{References}            
   \beginparmode
   \frenchspacing \parindent=0pt \leftskip=1truecm
   \parskip=8pt plus 3pt \everypar{\hangindent=\parindent}}
     
     
     
\gdef\journal#1, #2, #3, 1#4#5#6{               
    {\sl #1~}{\bf #2}, #3 (1#4#5#6)}            

\gdef\refa#1, #2, #3, #4, 1#5#6#7.{\noindent#1, #2 {\bf #3}, #4 (1#5#6#7).\rm} 

\gdef\refb#1, #2, #3, #4, 1#5#6#7.{\noindent#1 (1#5#6#7), #2 {\bf #3}, #4.\rm} 

\def\pr{\journal Phys.Rev., }

\def\endreferences{\body}

\def\endpage                    
  {\vfill\eject}
     
\def\endpaper                   
  {\endmode\vfill\supereject}

\def\ref#1{Ref.~#1}                     
\def\[#1]{[\cite{#1}]}
\def\cite#1{{#1}}
\def\(#1){(\call{#1})}
\def\call#1{{#1}}
\def\taghead#1{}
\def\frac#1#2{{#1 \over #2}}
\def\half{{\frac 12}}

\def\12{{1\over2}}

\def\la{\langle}
\def\ra{\rangle}
\def\ria{\rightarrow}

\def\x{{\bf x}}

\def\k{{\bf k}}
\def\q{{\bf q}}

\def\a{\alpha}

\def\ih{{i \over \hbar}}
\def\au{{\underline{\alpha}}}

\def\p{{\bf p }}

\centerline{\bf Decoherent Histories and the Emergent Classicality 
of Local Densities}

\author J.J.Halliwell 
\affil 
Theory Group 
Blackett Laboratory 
Imperial College 
London 
SW7 2BZ 
UK 
\vskip 0.5in 
\centerline {\rm Imperial Preprint/TP/98-99/53. Revised version. August, 1999} 
\vskip 0.1in  

\vskip 1.0in 

\abstract{In the context of the decoherent histories approach to
quantum theory, it is shown that a class of macroscopic
configurations consisting of histories of local densities (number,
momentum, energy) exhibit negligible interference. This follows from
the close connection of the local densities with the corresponding
exactly conserved (and so exactly decoherent) quantities, and also
from the observation that the eigenstates of local densities
(averaged over a sufficiently large volume) remain approximate
eigenstates under time evolution. The result is relevant to the
derivation of hydrodynamic equations using the decoherent histories
approach.}

\endtopmatter 
\endpage

The key step in many studies of emergent classicality from quantum 
mechanics is the demonstration that certain types of quantum states
of the system exhibit negligible interference.  Initial
superpositions of such states may therefore be effectively replaced
by statistical mixtures. This, loosely speaking, is decoherence, and
has principally been demonstrated for the situation in which
there is a distinguished system, such as a particle, coupled to its
surrounding environment [1].

Most generally, decoherence typically comes about when the variables
describing the entire system of interest naturally separate into
``slow'' and ``fast'', whether or not this separation corresponds
to, respectively, system and environment\footnote{$^{\dag}$}{See
Ref.[2] for a discussion of the  conditions under which
the total Hilbert space may be  written as a tensor product of
system and environment Hilbert spaces}. If the system consists of a
large collection of interacting identical particles, as in a fluid
for example, the natural set of slow variables are the local
densities: energy, momentum, number, charge {\it etc.}  These
variables, in fact, are also the variables which provide the most
complete description of the {\it classical} state of a fluid at a
macroscopic level. The most general demonstration of emergent
classicality therefore  consists of showing that, for a large
collection of interacting particles described microscopially by
quantum theory, the local densities become effectively classical.
Although one might argue that the system--environment mechanism
might play a role, since the collection of particles are coupled to
each other, decoherence comes about in these situations for a
different reason: it is because the local densities are almost
conserved if averaged over a sufficiently large volume
[3]. Hence, the approximate non-interference of
local densities is due to the fact that they are close to a set of
exactly conserved quantities, and exactly conserved quantities obey
superselection rules. 

Intuitively appealing though this argument is, it is clearly a {\it
quantitative} issue. The object of this letter is to show that,
under certain reasonable conditions,  local densities averaged over
a sufficiently large volume are indeed  approximately decoherent as
a result of their close connection to exact conservation.

We will approach the question using the decoherent histories
approach to quantum theory [3-5], which
has proved particularly useful for discussing emergent classicality
in a variety of contexts\footnote{$^*$}{The extent to which the
approach fully explains emergent classicality 
has been criticized [6]. This paper concerns
the mathematical properties of the approach, as it currently
stands, and adds nothing to that debate. See Ref.[7] and 
references therein for further discussion.}.
The central object of interest is the decoherence functional,
$$
D (\au, \au' ) =  {\rm Tr} \left( C_{ \au} | \Psi \ra \la | \Psi |
C_{\au'}^{\dag} \right)
\eqno(1)
$$
The histories are characterized by the initial state $ | \Psi \ra $
and by the time-ordered 
strings of projection operators $C_{\au} = P_{\a_n} (t_n) \cdots
P_{\a_1} (t_1) $
(where $\au$ denotes the string of alternatives $\a_1, \a_2 \cdots
\a_n$). Intuitively, the decoherence functional is a measure of the
interference between pairs of histories $\au$, $\au'$. When it is
zero for $\au \ne \au' $, we say that the histories are decoherent and
probabilities  $ p (\au ) = D (\au, \au ) $ obeying the usual 
probability sum rules may be assigned to them.  Although not
addressed here, one can then ask whether these
probabilities are strongly peaked about trajectories obeying
classical equations of motion. For the local densities, these
equations will be hydrodynamic equations, and these and closely
related aspects of emergent classicality have been pursued at
greater length elsewhere Refs.[2,7-9]. 

We consider the class of systems which are described at the
microscopic level by a Hamiltonian of the form
$$
H = \sum_j \left( { {\bf p}_j^2 \over 2 m } + \sum_{\ell >j } \phi
( | \q_j - \q_\ell | ) \right)
\eqno(2)
$$
For definiteness, we will concentrate on the case of a 
dilute gas with short-range interactions,
but it will be clear that the physical
ideas are reasonably general.
The local densities of interest are the number density
$n(\x)$, the momentum density ${\bf g}(\x)$ and the energy
density $ h (\x )$, defined by,
$$
\eqalignno{
n(\x) &= \sum_j \ \delta(\x -\q_j)
&(3) \cr
{\bf g}(\x) &= \sum_j \ {\bf p}_j \ \delta (\x- \q_j)
&(4) \cr
h(\x) &= \sum_j \ \left( { {\bf p}^2_j \over 2 m } + \sum_{\ell > j}
\phi  (| \q_j - \q_{\ell} |) \right)\ \delta(\x- \q_j)
&(5) \cr }
$$
(suitably ordered, in the quantum case).
We are interested in local densities smeared over a volume $V$.
The effect of this is to replace the delta functions with a window
function, denoted $\delta_V$, which is zero outside $V$ and $1$
inside. It is also 
useful to work with the Fourier transforms of the local densities,
denoted $n (\k) $, $g (\k) $, $h(\k )$. So, for example,
the local number density at wavelength $\k$ is
$$
n(\k) = \sum_j \ e^{i \k \cdot \q_j}
\eqno(6)
$$
Exact conservation is obtained in the limit $ k = | \k | \ria  0 $, 
or $V \ria \infty $ in (3)--(5).

We would like to compute the decoherence functional for histories
consisting of projections onto the operators (3)--(5). (The
construction of the projectors is described in more detail in
Ref.[7]). In the case of exact conservation, $k=0$,  we have exact
decoherence simply because the projectors in Eq.(1) all commute with
$H$ and with each other [10]. Our main task is therefore to show
that as $k$ increases from zero there is still a non-trivial regime
in which decoherence is approximately maintained. A significant
result of this type has been established already by Calzetta and Hu
for the case of local temperature $T(x)$ obeying  the diffusion
equation [9]. They took their initial state to be close to the
equilibrium state, whereas here, by contrast, initial macroscopic
superposition states are considered.

We begin by rewriting the exact conservation case in a simple way
that makes its generalization to locally conserved quantities more
apparent. Suppose the histories are projections onto some conserved
quantity, $Q$. Let the initial state be a superposition of
eigenstates of $Q$,
$$
| \Psi \ra = { 1 \over \sqrt{2} } \left( |a \ra + |b \ra \right)
\eqno(7)
$$
where $ \la a | b \ra = 0 $ and
$$
\hat Q | a \ra = a | a \ra, \quad \hat Q | b \ra = b | b \ra
\eqno(8)
$$
Since the $P_{\a}$'s are projections onto $Q$, $P_{\a}$
either annihilates or preserves
$ | a \ra $ and $ | b \ra $.
Take the case of a history with just two
moments of time (the generalization to more times is trivial).
The only non-zero off-diagonal terms of the decoherence functional 
are of the form
$$
\eqalignno{
D(\au, \au') 
&= \half {\rm Tr} \left( P_{\a_2} e^{ -\ih H t} | a \ra \la b | e^{
\ih H t } \right)
\cr
& = \half {\rm Tr} \left( P_{\a_2} | a_t \ra \la b_t | \right)
&(9)  \cr }
$$
But $Q$ is conserved, hence $[P_{\a_2},H] = 0 $ and
$$
\eqalignno{
P_{\a_2} | a_t \ra & = P_{\a_2 } e^{ - \ih H t } | a \ra 
\cr
& = e^{ - \ih H t } P_{\a_2} | a \ra = | a_t \ra
&(10) \cr }
$$ 
(or equals zero if $\a_2$ does not correspond to $a$).
It follows that
$$
\eqalignno{
D(\au, \au') & =
\half {\rm Tr} \left( P_{\a_2} | a_t \ra \la b_t  | \right) 
\cr
& = \la b_t | a_t \ra = \la b | a \ra = 0 
&(11) \cr} 
$$
and therefore we have decoherence.

Now suppose that the operator $Q$ is one of the local densities
(3)--(5), so is no longer exactly conserved. The steps up to Eq.(9)
still hold. But to go further, we need to know how the eigenstates
of the local densities behave under time evolution. A reasonable
supposition, which will be justified, is the following.
Let us suppose that under time evolution, the
eigenstates of $Q$ remain approximate eigenstates. That is, we
initially have (7), but under evolution to time $t$,
$ \hat Q  | a_t \ra \approx \la Q \ra | a_t \ra $
or, more precisely, 
$$
{ \left( \Delta Q \right)^2 \over \la Q  \ra^2 } << 1
\eqno(12)
$$
{\it i.e.}, the state remains strongly peaked in the variable $Q$
under time evolution. The states are then approximate eigenstates
of the projectors, so that in place of Eq.(10), we have 
the approximate result, $ P_{\a_2} | a_t \ra \approx | a_t \ra $
(or equals zero) as long as
the width of the projection is much
greater than the  uncertainty $ (\Delta Q)^2 $. Hence Eq.(11)
follows approximately, and we get approximate decoherence to the
extent that the approximation (12) holds.

The key point is therefore the following: approximate decoherence is
assured for histories of operators $Q$ whose eigenstates
have the property that they remain strongly
peaked in $Q$ under time evolution, as characterized by (12). To
demonstrate decoherence of the local densities, therefore, we need
only find their eigenstates, and show that they satisfy the
localization property (12) under time evolution. (Note, incidently,
that the above argument actually assures decoherence of {\it any}
variables $Q$ satisfying the localization property. The particular
significance of the local densities is that they are continuous
functions of the coarse graining scale $k$, so are guaranteed to
satisfy the requisite property if $k$ is sufficiently close to
zero.)

Since the three operators (3)--(5)
do not commute, exact simultaneous eigenstates cannot be found. However,
a useful class of approximate eigenstates of all three operators
are the states consisting of products of $N$ identical terms,
$$ 
| \Psi \ra = | \psi \ra \otimes | \psi \ra \otimes 
\cdots \otimes | \psi \ra
\eqno(13) 
$$ 
These may be shown to be eigenstates (of the
local number density, for example) by observing that
the object $ ( \Delta n (\x) )^2 /
\la n (\x) \ra^2 $ goes like $1/N$ for large
$N$ (see Ref.[7], for example).  It is essentially the
central limit theorem (see also Ref.[11]). For the number
and momentum density it relies on the fact that they are sums of
identical one-particle operators. For the local energy density, it
additionally requires the smearing volume to be sufficiently large,
compared to some lengthscale indicated by the interactions. Some
tuning of the state $ | \psi \ra $ can be carried out to ensure that
(13) is an optimal approximate eigenstate of all the  local
densities but this will not be done here. (Also,  the passage to
exact eigenstates of $n(\k )$, $g( \k ) $, $h ( \k ) $  as $k \ria 0
$ can be seen explicitly if the one-particle states  $ | \psi \ra$
are taken to be one-particle momentum eigenstates).

The question is now what happens to the eigenstates (13) of the local
densities under time evolution by the Hamiltonian (2). Consider
first the trivial but enlightening case in which there no
interactions. In this case, the time evolved eigenstates $ | a_t
\ra $ remain of the product form (13), so they are {\it still}
approximate eigenstates of the local densities (but with a
time-evolved eigenvalue) for the same reasons as above. 
Hence there is approximate decoherence.

Decoherence in the non-interacting case comes about for two reasons.
First, it is due to the fact that a state of the form (13) will
remain strongly peaked about the average values of the local
densities, $n(\x)$, $g( \x ) $, $h (\x ) $ under time evolution, and
thus the state is essentially undisturbed by the projectors (as long
as their widths are sufficiently large).  Secondly, it is due to the
almost trivial fact that the orthogonality of the two elements of
the initial state is preserved by unitary evolution. This second
fact is important because the first one is not always sufficient to
guarantee decoherence. Although the state remains strongly peaked
about the average values of the local densities, these average
values do not necessarily obey deterministic equations. In the case
of histories characterized by number density only, for example, $
\la n (\x) \ra $ at time $t$ is {\it not} uniquely determined by  $
\la n (\x ) \ra $ at the initial time (in the state (13)). That is,
in Eq.(9), $ | a_t \ra $ and   $ |b_t \ra $ may in fact be peaked
about {\it the same} value of number density, even though the
initial values are different.  The decoherence is therefore not in
fact due to an approximate  determinism (such as that used in the
phase space histories of Omn\`es [5]).  It is necessary
only that the evolved states are essentially undisturbed by the
projectors, and therefore that the two orthogonal components of the
initial state are eventually overlapped at the final time, as in
Eq.(11), to give zero.

The next and most important task is to show that the above story is
in fact still true, with qualifications, in the presence of
interactions. The complete description of $N$ interacting particles
is complicated but we can make some progress by making two
assumptions which are standard in kinetic theory [12].  It is
notationally convenient in what follows to work with a Wigner
function, rather than quantum state. Hence associated with the full
$N$--particle wave function is an $N$--particle Wigner function $
W_N (\p_1, \q_1, \cdots \p_N, \q_N ) $. Our first assumption is that
the  three--particle correlations are negligible.  It means that all
the physics is contained in the one and two--particle reduced Wigner
functions, $W_1 (\p_1, \q_1)$ and $ W_2 (\p_1, \q_1, \p_2, \q_2 ) $.

We again take as our initial state the
approximate eigenstate (13), and let it evolve, so correlations will develop.
The degree to which the particles become
correlated is contained in the two--particle distribution
$W_2$ of the evolved eigenstate. 
On general grounds, we expect that the inter-particle
correlations will only be important on some length scale $L$, and
beyond that length scale, they will be uncorrelated. That is,  we
will assume that
$$
W_2 (\p_1, \q_1, \p_2, \q_2 ) \approx W_1 (\p_1, \q_1 ) W_1 (\p_2,
\q_2 )
\eqno(14) 
$$
for $ | \q_2 - \q_1 | > L$, and otherwise $W_2$ will have a form
indicating non-trivial correlations. This is our second assumption.
It is physically reasonable for uncorrelated initial states of the
form (13) with a short range interaction 
(and it is in fact a key assumption in
the derivation of the Boltzmann equation [12]). It would not
of course be an appropriate assumption for correlated initial
states, but the point is that we are interested in approximate
eigenstates of the local densities, and a useful 
class of such states have the uncorrelated from (13).

Given the above assumptions, it is now reasonably straighforward to
argue that the state is still strongly peaked about the average values
of the local densities, as long as $V >> L^3 $. 
For example, for the number density, we have
$$
\la n(\x ) \ra = \sum_j \la \delta_V ( \q_j - \x ) \ra
= N \int_V d^3 \q \ p (\q )
\eqno(15) 
$$
where $ p ( \q ) $ is the one-particle 
probability distribution of $\q$ (obtained by integrating the
one-particle Wigner function over $\p $).
Similarly,
$$
\eqalignno{
\la n^2 (\x) \ra &= \sum_{j \ell} \ \la \delta_V (\q_j -\x ) \delta_V
(\q_{\ell} -\x ) \ra
\cr
&= N \la \delta_V \ra + (N^2 - N) \la \delta_V (\q_1 -x ) \delta_V
(\q_2 - x ) \ra
&(16) \cr}
$$
where we have used $\delta_V^2 = \delta_V $, and also an assumption
of identical particles to reduce the sum over $j,\ell$ to particles
labeled $1$ and $2$.
We now have
$$
\eqalignno{
( \Delta n (\x) )^2 
=& \la n^2 (\x)  \ra - \la n (\x) \ra^2 
\cr
= & N^2 \left( \la \delta_V (\q_1 - \x ) \delta_V (\q_2 - \x ) \ra
- \la \delta_V \ra^2 \right)
\cr
& + N \left( \la \delta_V \ra - \la \delta_V (\q_1 -\x ) \delta_V
(\q_2 - \x ) \ra \right)
&(17) \cr }
$$
If there is no correlation at all between the particles, the
coefficient
of $N^2$ would vanish, so $( \Delta n (\x) )^2 / \la n(\x )
\ra^2 $ would go like $ 1/ N $, which goes to zero as $N \ria
\infty$.
This is the standard central limit theorem result indicated earlier
for the non-interacting case. With interactions, the coefficient
of $N^2$ is no longer zero. We now need to show, therefore, that
this term
is still sufficiently small for  
$( \Delta n (\x) )^2 / \la n(\x ) \ra^2 $ 
to remain small as $N \ria \infty $. Introducing the
two-particle distribution $ p (\q_1, \q_2) $ (obtained by
integrating $\p_1, \p_2 $ out of $W_2$), it is readily shown that
the leftover terms as $N \ria \infty$ are
$$
{(\Delta n (\x) )^2 \over  \la n(\x )\ra^2 }
= { \int_V d^3 \q_1 \int_V d^3 \q_2 \left( p(\q_1, \q_2 )
- p(\q_1 ) p (\q_2 ) \right)
\over \left( \int_V d^3 \q \ p (\q ) \right)^2 }
\eqno(18)
$$
This is clearly zero if there are no correlations.
In the interacting case we use the assumption (14), which implies
that
$$
p(\q_1, \q_2 ) \approx p(\q_1 ) p (\q_2 )
\eqno(19)
$$
for $ | \q_1 - \q_2 | > L $, and otherwise non-trivial correlations
exist. Hence the integral in the numerator takes contributions
only from the region $ | \q_1 - \q_2 | < L $.

To see that (18) is small, note that in the numerator, the
integral is over a volume $V^2$ in the six-dimensional two particle
configuration space. If $ V << L^3 $, the factorization of  $ p
(\q_1, \q_2 ) $ for $ | \q_1 - \q_2 | > L $ makes no difference,
since $\q_1$ and $\q_2$ can never be far enough apart in the 
integrand (assuming $V$ is regular in shape). 
However, if $ V >> L^3 $, the $V^2$-sized integration
region is substantially reduced in size to $ V \times L^3 $. On dimensional
grounds the numerator is therefore proportional to a number of order $ V L^3
$, and the denominator to $V^2 $ (perhaps with other factors common
to both). This means that
$$
{(\Delta n (\x) )^2 \over  \la n(\x )\ra^2 } \sim { L^3 \over V }
\eqno(20) 
$$
This order of magnitude estimate becomes exact if we assume
that the probabilities are constant in the region of non-trivial
correlation (another common assumption of kinetic theory [12]).
Hence the state will be strongly peaked about the average of
$ n(\x )$ if $ V >> L^3$. 

It is possible to see on physical grounds why one expects a result
of the form (20) to hold quite generally. In the non-interacting
case we used the  central limit theorem result that $ ( \Delta n )^2
/ \la n \ra^2 $ goes like $1/N$.  In the interacting case, the state
is no longer of the product form (13), but an analagous result still
holds. The point is that the correlations that develop extend only
over a (typically small) volume of size $L^3$, so the system breaks
up into a large number of essentially identical uncorrelated regions
of this size. Therefore each smearing volume $V$, if much greater
than $L^3$, contains of order $V/L^3$ identical  uncorrelated
regions each of which contribute equally to the local density
averaged over $V$. Loosely speaking, a central limit theorem-type
result again applies, not to the $N$ uncorrelated particles in the
same state, but to the  $ V/L^3 $ uncorrelated regions. So $1/N$ is
replaced by $L^3 /V $ in the central limit theorem, and hence the
above result.

Similar results hold for the local momentum and energy density. We
have therefore demonstrated the desired result: a class of 
eigenstates of the coarse-grained local densities remain
approximate eigenstates under time evolution as long as the smearing
volume is much greater than the correlation volume of these states.
Decoherence of these variables then follows. More details of this
work, including a discussion of the approach to local equilibrium
and the emergence of hydrodynamic evolution equations,
will be published elsewhere [13].

\noindent{\bf Acknowledgements: }I am grateful to Todd Brun, Jim
Hartle, Ray Rivers and Tom Kibble for useful conversations.

\subhead{\bf References}

\def\pr{{\sl Phys.Rev.}}

\item{1.} E.Joos and H.D.Zeh, {\sl Z.Phys.} {\bf B59}, 223 (1985);
W.Zurek, in {\it Physical Origins of Time Asymmetry},
edited by  J.J.Halliwell, J.Perez-Mercader and W.Zurek (Cambridge
University Press, Cambridge, 1994); preprint quant-ph/9805065.

\item{2.} T.Brun and J.B.Hartle, quant-ph/9905079.

\item{3.} M.Gell-Mann and J.B.Hartle, in {\it Complexity, Entropy 
and the Physics of Information, SFI Studies in the Sciences of Complexity},
Vol. VIII, W. Zurek (ed.) (Addison Wesley, Reading, 1990);
{\sl Phys.Rev.} {\bf D47}, 3345 (1993).

\item{4.} R.B.Griffiths, {\sl J.Stat.Phys.} {\bf 36}, 219 (1984);
{\sl Phys.Rev.Lett.} {\bf 70}, 2201 (1993).

\item{5.} R. Omn\`es, {\sl J.Stat.Phys.} {\bf 53}, 893 (1988);
{\bf 53}, 933 (1988);
{\bf 53}, 957 (1988);
{\bf 57}, 357 (1989);
{\sl Ann.Phys.} {\bf 201}, 354 (1990);
{\sl Rev.Mod.Phys.} {\bf 64}, 339 (1992).

\item{6.} H.F.Dowker and A.Kent, {\sl J.Stat.Phys.} {\bf 82},
1575 (1996); {\sl Phys.Rev.Lett.} {\bf 75}, 3038 (1995);
A. Kent, {\sl Phys.Rev.} {\bf A54}, 4670 (1996).

\item{7.} J.J.Halliwell, {\sl Phys.Rev.} {\bf D58}, 105015 (1998).

\item{8.} C.Anastopoulos,
preprint gr-qc/9805074 (1998);
T.Brun and J.J.Halliwell, 
{\sl Phys.Rev.} {\bf 54}, 2899 (1996);
E. Calzetta and B. L. Hu, in {\it Directions in General
Relativity}, edited by B. L. Hu and T. A. Jacobson (Cambridge
University Press, Cambridge, 1993).

\item{9.} E.A.Calzetta and B.L.Hu, {\sl Phys.Rev.} {\bf D59},
065018 (1999).

\item{10.} J. B. Hartle, R. Laflamme and D. Marolf, 
\pr {\bf D51}, 7007 (1995).

\item{11.} J.B.Hartle, {\sl Am.J.Phys.} {\bf 36}, 704 (1968).

\item{12.} See for example, R.L.Liboff, {\it Introduction to the
Theory of Kinetic Equations} (Wiley, New York, 1969); K.Huang, {\it
Statistical Mechanics}, 2nd edition (New York, Chichester, Wiley,
1987).

\item{13.} J.J.Halliwell and J.B.Hartle (in preparation).

\endreferences

\end